\documentclass[longbibliography]{article}
\usepackage[letterpaper, margin=0.9in]{geometry}
\usepackage[utf8]{inputenc}

\usepackage{hyperref}
\hypersetup{colorlinks=true,urlcolor=blue,citecolor=blue,linkcolor=blue}
\usepackage{enumitem}        
\setlist{nosep}
\usepackage[super]{nth}
\usepackage{graphicx}
\usepackage{url} 

\usepackage{caption}
\usepackage{titling}
\newcommand{\subtitle}[1]{%
  \posttitle{%
    \par\end{center}
    \begin{center}\large#1\end{center}
    \vskip0.5em}%
}
\newcommand{\comment}[1]{}

\usepackage{array}
\newcommand{\PreserveBackslash}[1]{\let\temp=\\#1\let\\=\temp}
\newcolumntype{C}[1]{>{\PreserveBackslash\centering}m{#1}}
\newcolumntype{R}[1]{>{\PreserveBackslash\raggedleft}m{#1}}
\newcolumntype{L}[1]{>{\PreserveBackslash\raggedright}m{#1}}

\newcommand{\size}[1]{{\scriptsize #1}}
\newcommand{\intro}{\size{\index{Year!Intro}\textbf{Intro}}}
\newcommand{\bfy}{\size{\index{Year!BFY}\textbf{BFY}}}

\newcommand{\sm}{\size{\index{Class size!Small}\textbf{Small}}}
\newcommand{\me}{\size{\index{Class size!Medium}\textbf{Medium}}}
\newcommand{\lrg}{\size{\index{Class size!Large}\textbf{Large}}}

\newcommand{\electronics}{\size{\index{Content!Electronics}\textbf{Electronics}}}
\newcommand{\optics}{\size{\index{Content!Optics}\textbf{Optics}}}
\newcommand{\quantum}{\size{\index{Content!Quantum}\textbf{Quantum}}}

\newcommand{\eandm}{\size{\index{Content!Electromagnetism}\textbf{E\&M}}}
\newcommand{\waves}{\size{\index{Content!Waves}\textbf{Waves}}}

\newcommand{\notphyseng}{\size{\index{Majors!Not Physics nor Engineering}\textbf{NotPhysEng}}}
\newcommand{\physeng}{\size{\index{Majors!Physics or Engineering}\textbf{PhysEng}}}

\newcommand{\stem}{\size{\index{Majors!STEM}\textbf{STEM}}}

\newcommand{\other}{\size{\index{Majors!Other}\textbf{Other}}}

\usepackage{makeidx}
\makeindex

\title{\textbf{Teaching labs during a pandemic:\\ Lessons from Spring 2020 and an outlook for the future}}

\author{Michael F. J. Fox, Alexandra Werth, Jessica R. Hoehn, and H. J. Lewandowski \\ 
{\footnotesize Department of Physics, University of Colorado, Boulder, Colorado 80309, USA}\\
{\footnotesize JILA, National Institute of Standards and Technology and University of Colorado, Boulder, Colorado 80309, USA} 
}
\date{July 2020}

\begin{document}
\maketitle
\begin{abstract}
We report results from a survey of lab instructors on how they adapted their courses in the transition to emergency remote teaching due to the COVID-19 pandemic. The purpose of this report is to share the experiences of instructors in order to prepare for future remote teaching of labs. We include summaries of responses to help illustrate the types of lab activities that were done, learning goals for the remote labs, motivations for instructors' choices, challenges instructors faced, and ways in which instructors and students communicated. This is a first step in a larger project as part of an NSF RAPID grant to understand what happened during the switch to remote labs and how it impacted teaching methods and student learning. 

\end{abstract}
\newpage

\tableofcontents

\newpage
\section{Introduction}
In the spring of 2020, due to the COVID-19 pandemic, colleges and universities across the world rapidly transitioned classes and activities to be conducted remotely. This transition presented particular challenges for laboratory courses. This report forms part of a larger project studying the impact of public health restrictions on teaching methods and student learning in physics laboratory courses at the undergraduate level. The motivation for this report is to provide feedback, resources, and ideas to the community of physics instructors, detailing what instructors did and what worked well, before Fall 2020 classes begin. This report is distinct from other online recommendations developed for teaching remote labs, such as PhysPort \cite{physport} or ALPhA \cite{alpha}, in that the ideas come from the experiences of a large range of instructors and students. The nature of this report is a presentation and organization of collected data, rather than an analysis of a research question. A full analysis for a peer-reviewed publication will occur later. 

We define remote labs to encompass any continued instruction of a course that was considered a lab course prior to the rapid transition to remote work, in which the instructor and all students were no longer present at the same location. The data in this report primarily come from: (1) a survey sent out to lab instructors (the instructor survey) on April \nth{30} 2020, with the majority of responses from 106 instructors being received within the following 2 weeks, and (2) a supplementary survey appended to the standard E-CLASS~\cite{Zwickl2014} assessment administered to over 2600 students in over 50 courses (the student survey). The instructor survey contained both closed- and open-response questions that asked instructors about their experience transitioning to remote lab instruction. The student survey also included both closed- and open-response questions; however, here, we report only some data from the closed responses on the student survey to supplement the responses to the instructor survey. In some areas of the report, we provide examples from an ongoing interview study in which we are interviewing a handful of instructors to gain a more in-depth understanding of their approach to remote lab teaching. 

We report the quantitative results from the closed-response questions in the instructor survey in order to illustrate general trends, as well as variations between instructors' approaches to the challenge set before them. We support the quantitative data with examples (quotes) from the open-response questions to provide exemplars of approaches taken by instructors and the ways in which they were successful. These examples come from a wide range of different instructional environments---first-year introductory courses to graduate labs; various class sizes (from less than 10 to 100s of students); courses for non-scientists to courses for physics majors; and from community colleges to research intensive institutions. While each of these contexts have their own unique challenges, and there is clearly not a one-size-fits-all solution, we hope that, by illustrating a range of what worked well, instructors can draw inspiration from others in the community. In determining what worked ``well,'' there are a variety of metrics of success that instructors bring to bear, which is informed by their individual contexts, values, teaching approaches, and goals. Success of given strategy or course may be measured by: equitable implementation (i.e., do all students have access to the same learning opportunities?), student learning outcomes, student affect (i.e., did students enjoy the course?), addressing learning goals of the course (whether preserved from the in-person course or novel to remote teaching), ease of implementation for the instructor, or simply making it through the term. 

We structure the report around a number of themes that we consider to be important, and that lab instructors often consider, when thinking about the design and implementation of a course. These are: Section~\ref{sec:motivations-and-challenges}: Motivations of, and challenges faced by, lab instructors, Section~\ref{sec:learning-goals}: Learning goals, Section~\ref{sec:lab-structure}: Lab activities, Section~\ref{sec:student-agency}: Student agency and engagement, and Section~\ref{sec:communication}: Communication. The topic of Section~\ref{sec:motivations-and-challenges} provides an outline of the unique situations lab instructors found themselves in during Spring 2020. The following section on learning goals acts as an overview of what instructors did, as many of the choices made in subsequent sections depend upon the learning goals for any particular course. Within each subsequent section, we discuss aspects of the technologies used, and challenges faced, by instructors, as well as linking back to the learning goals of Section~\ref{sec:learning-goals}. We conclude in Section~\ref{sec:next-semester} with a discussion and recommendations for physics labs going forward in a remote or hybrid (remote and in-person) fashion. We also provide an index for instructors that wish to identify particular examples of resources related to the subject of their lab course, such as electronics or optics. Finally, in Appendix~\ref{sec:technology}, we include tables of technological resources that instructors reported using in their remote lab courses. Before presenting the results, we provide, in the following section, a summary of the sample of instructors who completed the survey.

\subsection{Survey sample}

The instructor survey was completed for 129 courses by 106 unique instructors. A majority of the respondents came from 4-year colleges (55\%). Approximately 8\% of the responses were from classes at 2-year colleges, 5\% from Master's granting institutions, and 32\% from PhD granting institutions. 61\% of courses were first year (introductory) labs and 39\% were beyond first year labs. Approximately 30\% of the labs were taught to primarily non- physics or engineering majors, 60\% were taught to primarily physics and engineering majors, and 10\% mixture of majors. Most respondents switched to remote teaching part way through the term, though 17\% of respondents were remote for the entire term (typically from quarter/trimester systems).

\subsection{How to navigate this report}
In order to facilitate the extraction of relevant and useful information from this report, we have labeled each example with at least 3 tags. These tags identify the context of an example and are intended to help the reader assess whether such an activity or approach would have similar effectiveness in their own situation. The page locations of each tag are provided in the Index.

The first label describes whether the course is at the introductory level (\size{\textbf{Intro}}), or is beyond the first year (\size{\textbf{BFY}}). The second label describes the majority of students who enroll in the course, based on their major: Physics and Engineering majors (\size{\textbf{PhysEng}}); STEM majors (\size{\textbf{STEM}}) i.e., including physics and engineering; not Physics nor Engineering majors (\size{\textbf{NotPhysEng}}); non-Science majors (\size{\textbf{Non-science}}); mainly Physics (\size{\textbf{Phys}}); mainly Math (\size{\textbf{Math}}); and other/non-classified (\size{\textbf{Other}}). The third label describes the size of the class. Classes with less than 25 students are labeled (\size{\textbf{Small}}); classes with between 25 and 100 students inclusive are labeled (\size{\textbf{Medium}}); classes with over 100 students are labeled (\size{\textbf{Large}}).

In addition to this labeling, we have included an index at the end of the report, so that the reader may quickly navigate to specific examples of interest. Quotes with information relevant to various physics subject matter are additionally labeled, and indexed as such. These content labels are: 
[\size{\textbf{Mechanics}}, 
\size{\textbf{E\&M}}, 
\size{\textbf{Waves}},
\size{\textbf{Electronics}}, 
\size{\textbf{Optics}}, 
\size{\textbf{Quantum}}, 
\size{\textbf{Astro}}].

\section{Motivations of, and challenges faced by, lab instructors}\label{sec:motivations-and-challenges}
\begin{figure}[ht]
    \centering
    \includegraphics[scale=0.6]{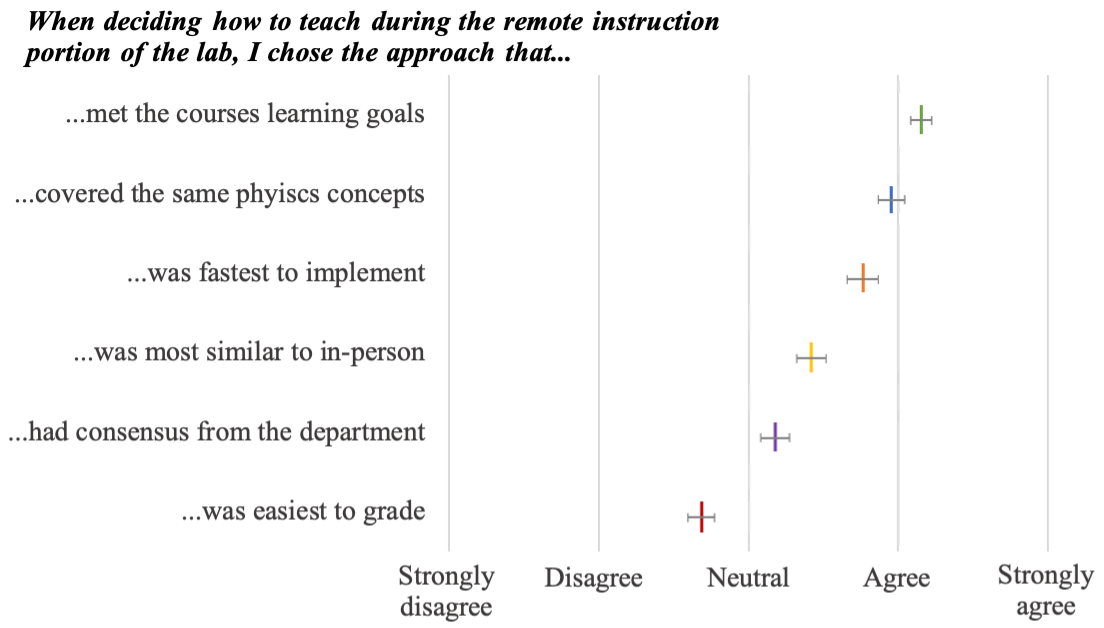}
    \caption{Instructors were asked to ``Rank how much you agree with the following statements.'' We show the mean response from 121 survey responses and the error, which represents one standard error of the mean. We calculated the mean by assigning a response of ``Strongly disagree" = 0, ``Disagree" = 1, and ``Neutral" = 2, ``Agree = 3", and ``Strongly agree = 4".}
    \label{fig:motivation}
\end{figure}

We begin by examining the motivations for, and challenges of, transitioning to remote lab instruction as expressed by the instructors who completed the instructor survey. We found that, although the motivations varied across the group of instructors, most people were driven by meeting the course learning goals and covering the same content as before the transition to remote instruction (see Figure \ref{fig:motivation}). While grading and having departmental consensus often represented constraints for instructors, these were not the primary motivators when designing the remote version of the course. Another motivation that was not represented in the closed response questions, but that we saw multiple times in the open responses was ensuring the remote course was equitable---i.e., all students in the class had access to the resources they needed to learn and thrive. For example, one instructor explained they \textit{``had to find things that worked that students could do without buying stuff.''} [\intro, \physeng, \lrg] For another, their main motivation was to ensure the well-being of their students: \emph{``I prioritized mental health by holding mental health check ins at the beginning of every class period.  This really helped the class to create a community and also re-enforced with the students that I valued them as people first.  I have found that students will work harder and learn more if you care for them as a whole person."} [\intro, \notphyseng, \sm]

\begin{figure}[ht]
    \centering
    \includegraphics[scale=0.6]{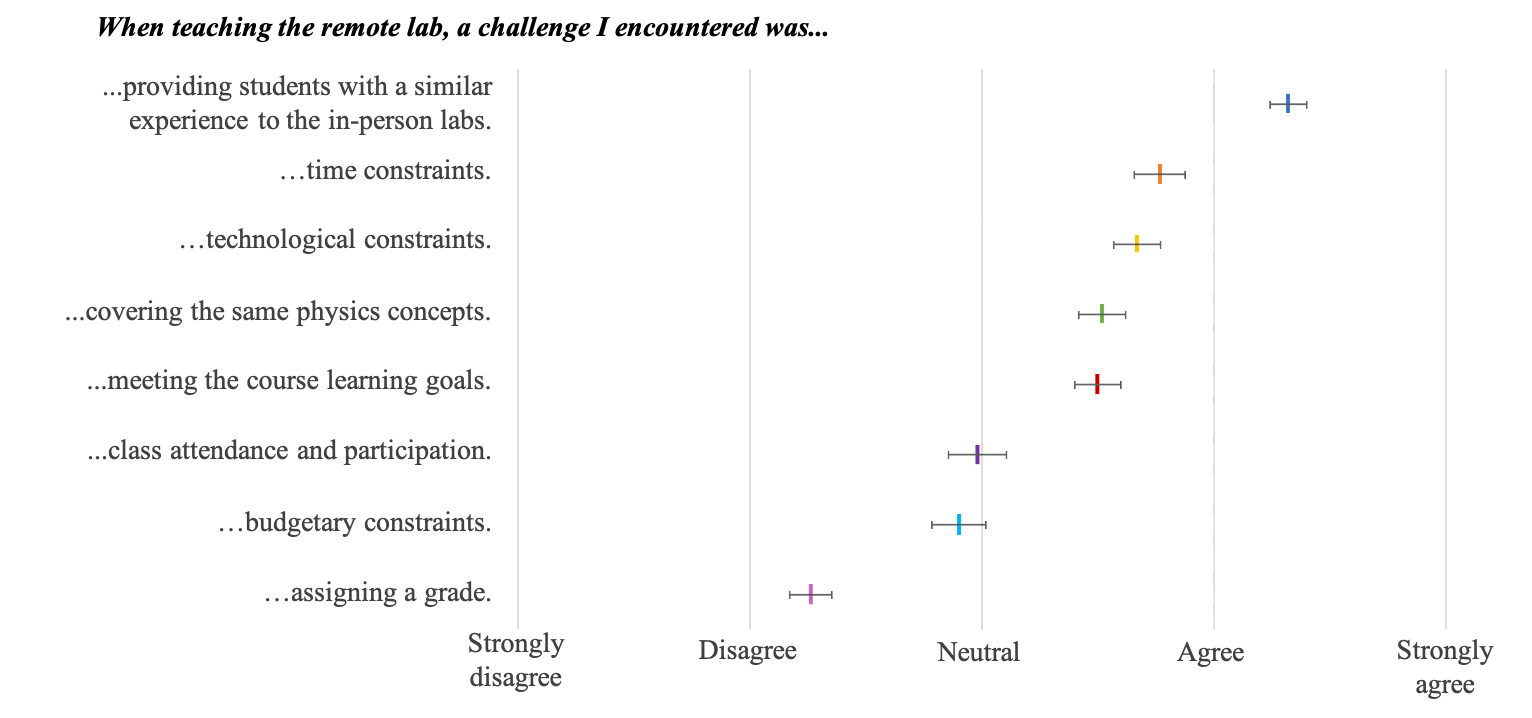}
    \caption{Instructors were asked to ``Rank how much you agree with the following statements.'' We show the mean response from 111 survey responses and the error which represents one standard error of the mean. We calculated the mean by assigning a response of ``Strongly disagree" = 0, ``Disagree" = 1, and ``Neutral" = 2, ``Agree = 3", and ``Strongly agree = 4".}
    \label{fig:challenges}
\end{figure}

Additionally, we asked instructors to rank each challenge they faced during this transition on a Likert scale. The most common reported challenge instructors faced was making the remote class as similar to the in-person version as possible. Instructors also cited time and technology constraints as major challenges. Grading did not seem to be a problem for too many people, perhaps because a large number of institutions switched to pass/fail grading schemes, or because many instructors were encouraged to be more lenient with their grading in the remote situation. Responses to the statements on class attendance/participation and budget were somewhat polarized (which is not represented by the mean shown in Figure \ref{fig:challenges}). Other challenges that were seen in the open-responses were personal factors for the instructor (e.g., family responsibilities), student engagement, group work, and equity for the students. For example, one instructor said, \textit{``I could imagine a class where experiments are done by the students at home, but given the different life circumstances of students, the class would likely not be an equitable experience.''} [\bfy, \physeng, \sm, \quantum] Another had challenges using simulations that used Java instead of HTML5 and expressed that the biggest challenge they faced was \emph{``choosing simulations all students can use on different hardware.''} [\intro, \other, \me] Challenges with group work were not only expressed by the instructors, but it was also one of the biggest challenges expressed by the students. 

\begin{figure}[ht]
    \centering
    \includegraphics[scale=0.6]{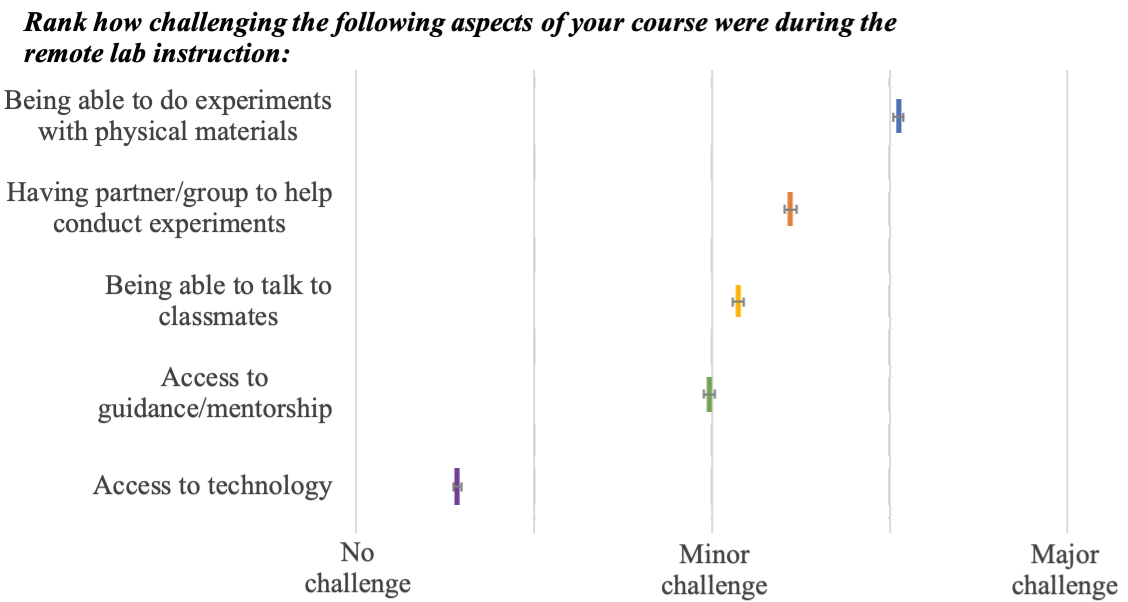}
    \caption{Students were asked to ``Rank how challenging the following aspects of your course were during the remote lab instruction.'' Students could choose either ``No challenge", ``Minor challenge", or ``Major Challenge".  We show the mean response from 2260 students and the error bars represent the standard error of the mean. We calculated the mean by assigning a response of ``No challenge" = 0, ``Minor challenge" = 1, and ``Major challenge" = 2.}
    \label{fig:studentchallenges}
\end{figure}

In addition to the instructor survey, we administered supplemental questions with the E-CLASS \cite{Zwickl2014}.  The most common major challenge that students reported was not being able to do experiments with physical materials  (Figure \ref{fig:studentchallenges}). The second most common challenge (on average) was not ``having a partner/group to help conduct experiments". While the majority (75.6\%) of students reported not facing a challenge associated with access to technology, 545 students reported access to technology as a minor challenge and 104 students reported it to be a major challenge. Additionally, the survey was administered via the internet so these numbers are likely underestimating the more severe cases of lack of access to technology. In order to ensure that lab (and all) classes are equitable, we recommend recognizing and addressing students' challenges and access to technology in current and future remote/hybrid course design.

Despite these myriad challenges, physics lab instructors rose to the occasion and employed a variety of creative approaches and strategies in order to provide opportunities for students to access ``lab-like'' learning online. As part of this report, we hope to provide examples and recommendations of ways to create productive remote lab experiences and collaborations. We will focus on the two primary motivating factors---meeting course learning goals and covering the same physics concepts---while acknowledging and incorporating potential solutions that will be equitable and as easy as possible to implement. Of course, we note that many of these solutions and outcomes are highly dependent on specific contexts (class size, student population, individual student and/or instructor circumstances, etc.) and hope to provide instructors with a wide variety of options that they may consider in the context of their own situation.

\section{Learning goals}\label{sec:learning-goals}
 
\begin{figure}[h!]
    \centering
    \includegraphics[scale=0.9]{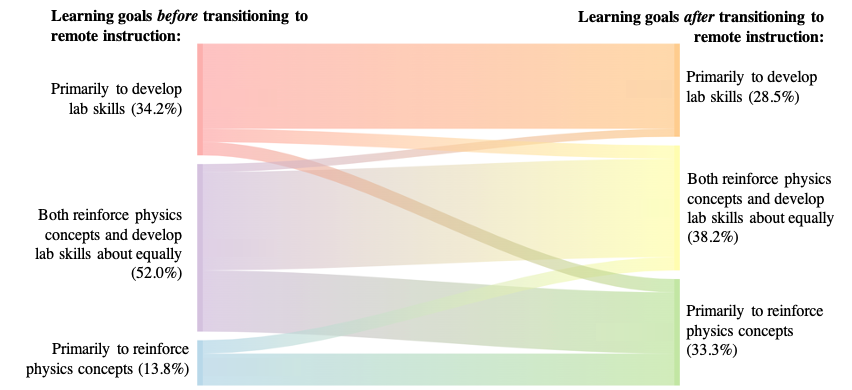}
    \caption{The Sankey plot shows the change in learning goals of the instructors who completed the instructor survey from before (left side of plot to after (right side of plot) remote instruction. The lines represent the direction of change from before to after and the width of the line is proportional to the number of instructors who reported that type of transition. }
    \label{fig:LearningGoals}
\end{figure}

While there exists a wide range of different implicit and explicit learning goals for labs that vary depending on institute and course, the physics education research literature generally categorizes these goals into two groups: either developing experimental skills or reinforcing physics concepts~\cite{Wilcox2017, Holmes2018}.

After the transition, the courses were approximately evenly distributed across learning goals that focused primarily on concepts, primarily on skills, and both concepts and skills equally. Many instructors shifted their learning goals to be focused primarily on reinforcing concepts during the remote version of the course; this shift came principally from people who originally had learning goals focused on \textit{both} concepts and skills (Figure~\ref{fig:LearningGoals}). This aligns with the literature, which finds that many proponents of online labs value learning physics concepts (i.e., content and theory) where proponents of hands-on labs often value design skills and collaborative skills~\cite{Ma,foreman,Corter,brinson}.

We believe pivoting learning goals of a lab course to focus more on concepts, given the extenuating circumstances, may have been a reasonable, productive, and effective solution. However, as we see from the instructor survey, the majority of courses with primary learning goals associated with skills maintained those learning goals after the transition, with many people finding creative ways to focus on laboratory skills in the remote classes. One survey respondent said, \textit{``We took this as an opportunity to completely redefine the goals of the course and try some ideas that likely would not have been seriously considered during a normal quarter.''} [\bfy, \physeng, \me] Whether trying to just survive the transition to remote instruction or using it as an opportunity to transform the course, instructors employed a variety of strategies to address their learning goals. In the following sections, we provide a few examples of ways to conduct remote labs focusing on lab-skill learning goals and some examples of ways to conduct remote labs focusing on concept learning goals.

\subsection{Skills}
The ability to maintain a focus on experimental skills during remote instruction depends on the resources available to students, as well as on what skills are considered important. In this section, we describe some common approaches to maintaining the development of skills as a learning goal in remote lab courses.

{\bf Hands-on learning from home.} The obvious challenge faced by remote instruction is the potential absence of hands-on interaction with measurement devices and experimental apparatus. This was more of a concern for advanced labs than intro-level labs, where more sophisticated and expensive equipment is usually used. For intro-labs, there were two common approaches: (1) to send equipment home to students---\textit{``The last three weeks were done with lab kits that I mailed to them in advance made almost entirely from materials that I already had in lab.''}[\intro, \notphyseng, \sm], and (2) to get students to use resources they had at home, with many instructors taking advantage of the prevalence of smart phone ownership among their students (while being aware that not all students would necessarily have access to these tools)---\textit{``I was able to incorporate a measurement that was consistent with the learning goals of the last two labs. Luckily they were optics... I had them measure the focal length of their cell phone camera lens based on the recent paper~\cite{Antoine2020}. Worked well!''} [\intro, \notphyseng, \me, \optics] In Section~\ref{sec:student-collected-data-at-home}, we discuss the range and types of hands-on activities students engaged in at home. Of course, these solutions may also be applicable to advanced-level lab courses depending on their specific aims. For example, open-ended project work may be more flexible with the types of equipment that students are expected to use, provided that appropriate methods of measurement and analysis are applied to answer the research questions posed.

{\bf Simulations.} Some lab courses switched to using simulations as sources of data collection and measurement: \textit{``Given the original design of the lab activities, a combination of Fritzing and Multisim Live allowed students to practice many of the skills I had already planned to address.''} [\bfy, \physeng, \sm, \electronics] While simple simulations may not be able to replicate the troubleshooting aspect of performing experiments in real life, the example above of using Fritzing may emulate more what working on circuit design is like for professionals, as it allows for the design and testing of circuit boards with the production of plans that could be sent to be manufactured. More on using simulations will be discussed in Section~\ref{sec:collect-data-from-sims}.

{\bf Provide the data.} A common learning goal for labs includes skills associated with data and uncertainty analysis. The development of these skills does not necessarily require students to collect their own data, though an understanding of how the data was measured and how it should be interpreted may be diminished. Therefore, many instructors sent data that they had collected, generated, or uncovered from previous students' work, which we discuss in Section~\ref{sec:instructor-provided-data}. Alternatively, instructors asked students to review data from scientific publications or publicly available data sets. 

A number of courses included a proposal writing or experimental design aspect (even before transition); see Section~\ref{sec:written-comms} for more details about writing in the remote-lab environment. Having students propose or design experiments can continue in the remote context, even if students do not have access to the necessary lab equipment to actually carry out the experiment. In one case, an instructor of an advanced lab took the following approach: \textit{``Student groups developed data collection plans to use with equipment they were already familiar with. Instructors then collected data according to student plans.''} [\bfy, \physeng, \me] This is an example of how some instructors tried to replicate the in-lab experience of student ownership of data~\cite{Dounas-Frazer2017,Dounas-Frazer2019}. This recurring theme of student agency is discussed in Section~\ref{sec:student-agency}.

{\bf Science communication as a skill.} Courses where broader skills-based learning goals were dominant were less affected by the transition to remote instruction. For example, courses focusing on the development of communication skills (see Section~\ref{sec:communication}) could still get students to produce written work and provide feedback to them. As, in most situations, students had gathered some data already, this led to an opportunity to highlight the value associated with making good lab notes~\cite{Stanley2018}: \textit{``Even though no lab work occurred after remote instruction began, students had to rely on their notebooks and previous data collection to complete required oral presentations and written reports, both considered part of `lab skills.' (i.e., experimental physics skills)''} [\bfy, \physeng, \sm]. For more discussion on how students used data they had previously gathered see Section~\ref{sec:previously-gathered-data}. Some instructors also mentioned that, today, online collaboration is a realistic scientific practice, and thus they wanted their students to be able to develop that skill during the remote lab course. 

{\bf Investigative science learning environment (ISLE).} One instructor, who was teaching an intro class with combined lecture and lab components, found that remote ISLE-like activities~\cite{ISLEs3, ISLEs2, ISLEs1} were more effective than recorded video lectures. They noted: \emph{``My lectures which had been productive during the term were largely ineffective in the new setting... I ascribe this to the more passive nature of viewing video... I mention this because it has made me rely much more on ISLE like activities."} Students had the opportunity to interact with demos using household materials such as investigating \emph{``static electricity with sticky tape"}, through live demonstrations by the instructor, and activities where students \emph{``guided [the] instructor...during live video conference in the conduct of the experiment and used data collected then together with video analysis of the experiment clips made during the session."} A class taking an ISLE approach may not only help with ``Zoom fatigue" by creating a more interactive class, but also provide students with the many other benefits that ISLEs enables, such as constructing physics knowledge by engaging in inquiry cycles that replicate the approach used by physicists to construct knowledge~\cite{ISLEs1}.

\subsection{Concepts}
Labs have often been used as a way for students to see in action the physical phenomena they have been studying in lecture/theory courses. It may be argued that the learning goal of reinforcing students' understanding of physics concepts does not need to rely as much on hands-on experience, as do goals associated with developing skills. In this section, we describe some common or interesting approaches taken by instructors in our sample who were teaching courses with learning goals associated with developing student understanding of physics concepts.

{\bf Video demonstrations.} The exposure to the act of performing measurements through videos, both videos made by the instructor or publicly available (e.g., YouTube), was found to be valuable for teaching concepts. One instructor explained, \textit{``The lab videos showing the data being taken went very well, and students reported that they understood the concepts better by seeing what the apparatus looked like and what kind of measurements could actually be done.''} [\intro, \notphyseng, \me, \optics] More on instructor made videos will be discussed in Section~\ref{sec:instructor-made-videos}.

{\bf Simulations.} As documented in previous research~\cite{Jimoyiannis2001, Perkins2006, Adams2010}, simulations were found to be very useful for reinforcing physics concepts: \textit{``Since the goal was primarily to explore physics concepts, I think the use of simulations helped us to still meet that goal.''} [\intro, \notphyseng, \me] This is particularly true as some simulations have been developed to address specific and common student difficulties~\cite{Guangtian2011}. More on simulations will be discussed in Section~\ref{sec:collect-data-from-sims}.

\comment{{\bf Sending equipment to students.} Some instructors managed to send equipment to students for courses with reinforcing physics concepts as a learning goal: \textit{``Students understand physics [concepts] by manipulating everyday objects much better than I expected.''} [\intro, \notphyseng, \sm] This approach has logistical and budgetary challenges, especially for larger introductory courses. There was, however, a general sense from instructors that finding some way to give students a hands-on experience was an essential part of the laboratory experience. This will be discussed further in Section~\ref{sec:student-collected-data-at-home}.}

\section{Lab activities}\label{sec:lab-structure}

There were a variety of approaches taken when transitioning to remote labs, with the most common being: providing students with data to analyze; conducting lab activities via simulations; having students watch videos of the instructor or TA conducting the lab; and completing experiments at home with household equipment or equipment sent by the instructor. In this section, we discuss all 7 of the main types of activities used in order of the most frequently reported on the instructor survey. We end the section with a discussion of how instructors used writing as an important element of remote lab classes.  

\subsection{Instructor provided data}\label{sec:instructor-provided-data}
In place of students collecting their own data, many instructors provided data to students. These data sets were sourced in a variety of different ways, where the instructor:
\begin{enumerate}
    \item completed the experiment and sent a data set to students;
    \item sent students copies of the lab notebooks of students from previous years;
    \item provided data from a published paper for students to (re-)analyze;
    \item provided access to open-source data (e.g., COVID-19 data).
\end{enumerate}
The efficacy of providing data to students instead of students collecting the data themselves depends on what the learning goals of the course are. The interested reader may find some more discussion of this in Priemer \textit{et al.} 2020~\cite{Priemer2020}. We describe below in more detail some examples of how this kind of activity may work.

{\bf Instructor provides data they collected from an experiment.} An interesting example where an instructor provided data to students to analyze is where the instructor \textit{``tried to provide more videos (and in some cases data) than necessary to...give students the opportunity to choose which pieces they would use.''}[\intro, \notphyseng, \me] This choice was deliberate in order to encourage students to \textit{``make their own judgment calls''} similar to the decision making process students would face in in-person labs. One thing to keep in mind when implementing such an activity is to communicate the expectations of what to do with the data so that students are not \textit{``overwhelmed because they [think] that they needed to use it all.''}

{\bf Analysis of open-source data.}\label{sec:big-data} Another option is to provide students with big data and/or data from an active research experiment. NASA \cite{nasa.github.io}, CERN \cite{cern}, and LIGO \cite{gw} all have open source data available to the public and there are plenty of publicly available data sources (e.g., meteorological, air pollution, and astronomy data). For example, one instructor \emph{``did a data analysis/modeling lab where students used publicly available COVID-19 data to make plots and develop their own growth models. This was well received and helped students feel like they were doing something relevant and meaningful.''} [\intro, \physeng, \me] However, this type of data often requires some experience, expertise, and time to access it and prepare it to be suitable for students to handle. CERN and LIGO provide some tutorials and software on their websites to get started. Alternatively, instructors could use data from their own or a colleagues research. Working with local experimental data not only provides students with an authentic, research-like experience, but could potentially be beneficial for the research as well if taught as a course-based research experience (CURE) \cite{CURE1, CURE2}. 

\subsection{Collect data from simulations} \label{sec:collect-data-from-sims}
Simulations allow students to interact with models of physical phenomena via their computers or smartphones. The complexity of these models corresponds more or less with how well they are able to emulate hands-on labs. For some purposes, simpler simulations that highlight only the phenomenon of interest can be more effective at achieving certain learning objectives. Conversely, more complicated simulations, with larger parameter spaces to explore, could engage students with decision making and troubleshooting learning goals of some lab courses.

Many instructors turned to readily available simulations to conduct their labs when transitioning to remote setting. The simulations that were most useful were those that: 
\begin{enumerate}
    \item allowed students to gather data: \emph{``students acquired data by changing an independent variable in the PhET simulations''} [\intro, \notphyseng, \me];
    \item had structured materials around the simulations, such as lab guides.
\end{enumerate}
Some instructors mentioned that the simulation labs were so successful that they plan to continue using simulations when back in person, as pre-lab or supplemental activities: \emph{``I might use them as part of a class even with in-person learning.''} [\bfy, \physeng, \sm]. The most commonly reported set of simulations used were those produced by \href{https://phet.colorado.edu/}{PhET}, though many other providers of simulations were also reported, such as those associated with textbooks (\href{https://matterandinteractions.org/student/}{Matter \& Interactions}, and \href{http://www.physics.pomona.edu/sixideas/resources.html}{Six Ideas That Shaped Physics}). Due to the quick turn around needed in the transitions to remote labs, many instructors took advantage of commercial simulations with packaged teaching resources: \emph{``I found the \href{https://virtualphysicslabs.ket.org/}{KET} simulations and curriculum to be useful as an emergency solution,''} [\intro, \stem, \me]. A full list of reported simulation resources can be found in Table~\ref{tab:simulations}.

Many electronics labs found simulations particularly useful because they were able to use software like \href{http://bwrcs.eecs.berkeley.edu/Classes/IcBook/SPICE/}{SPICE}, MATLAB's Simulink and \href{https://www.mathworks.com/products/simscape.html}{Simscape}, \href{https://fritzing.org/home/}{Fritzing} or \href{https://www.multisim.com/}{Multisim Live} to build and model `real' circuits. The fact that these tools are used in industry also meant that students could still have an authentic lab experience.

A number of instructors used the commercial web application \href{https://www.pivotinteractives.com/}{Pivot Interactives}: ``The two labs that I set up on Pivot Interactives worked really well.'' [\intro, \notphyseng, \sm] The application is a hybrid of simulation and video analysis, where real experiments have been filmed with a variety of different parameter selections. It allows the student to explore the real-world parameter space and, using overlaid measurement tools, perform measurements from the videos. Additionally, each simulation has associated online questions and resources.

\subsection{Students watched a video of the instructor doing the lab}\label{sec:instructor-made-videos}

Many instructors said that they utilized videos of themselves, or teaching assistants (TAs), conducting the lab. These videos could be shown synchronously or asynchronously and had a number of different purposes, such as:
\begin{enumerate}
    \item an introduction to the lab;
    \item context for data to be analyzed;
    \item a means for students to record measurements;
    \item and an opportunity for students to direct the instructor in doing the experiment.
\end{enumerate}

The results of the instructor survey expressed a variety of different approaches to these videos, as well as a variety of degrees of success. For example, one instructor felt that \emph{``abstract concepts like diffraction from a single slit did not make sense until they saw the video and worked with the numbers.''} [\intro, \notphyseng, \me, \optics] Another was impressed with their students' trouble-shooting skills when \emph{``no guidance for accounting for [camera parallax] was provided, and yet all groups accounted for it or scaled their video in a way that it would not affect the data.''} [\intro, \physeng, \sm, \eandm]. These anecdotal findings correspond with some of the literature, for example Kestin \textit{et al.} (2020)~\cite{Kestin2020} found that video demonstrations are more effective learning tools than live demonstrations and that students reported the same level of enjoyment from both.

In contrast, a class \emph{``used a combination of PhET simulations and analysis of canned data after watching a video of the data collection"} and found that the \href{https://phet.colorado.edu/}{PhET} simulations were \emph{``much [more] effective and useful to the students than the [videos]."} [\intro, \notphyseng, \me] Although seemingly straightforward, creating an edited and professional looking video can take a surprisingly long time: \emph{``more than 1 hour for a 7 minute video''} [\bfy, \physeng, \sm, \quantum] and this often constrained instructors' use of recorded videos. Additionally, one has to be aware of how videos may not be suitable for students with cognitive or physical disabilities: \emph{``Many other faculty are using recorded videos of experiments--I choose not to because I do not think these videos are... accessible''} [\intro, \physeng, \sm]. 

One concern among instructors was whether students were watching asynchronous videos (see Section~\ref{sec:asynchronous}). A number of ways of addressing this concern were reported: one was to have students complete a set of questions after having watched the video on its content. Another used \href{https://go.playposit.com/}{PlayPosit} software to embed questions into the video as part of the pre-lab \cite{playposit}.

\subsection{Student collected own data at home}\label{sec:student-collected-data-at-home}

Maintaining a hands-on experience was commonly reported as a major motivation for choices made when moving to remote classes. Some instructors canceled the remainder of their classes because this was not feasible (due to time, budget, institutional, personal, or other constraints). Other instructors, who did not manage to incorporate a hands-on element in their lab course during the rapid transition to remote learning reported that they plan on including some aspects in the following semester. There were two main approaches to students collecting their own data at home: (1) to use household equipment; and (2) for the instructor to send equipment to students.

{\bf Using household equipment.} Using household equipment can be a fast, easy, and effective way for students to have a hands-on experience while being remote. However, it is important to recognize the issue of student equity. For example, most college students will have access to a smart phone and a computer, but there are still many who do not---especially when they leave campus (see discussion of challenges in Section~\ref{sec:motivations-and-challenges}). We recommend surveying the students before choosing this approach and having regular check-ins throughout the semester. One instructor said \textit{``I wish I had an inventory of technology that students had at home so I could have been better prepared to help troubleshoot or find alternate programs for data analysis and maybe felt less restricted in terms of not doing an experimental project.''} [\bfy, \physeng, \sm] Even access to simpler materials, like tape and magnets, proved to be an issue for some students. When implementing lab activities in which students are expected to use household equipment, we recommend ensuring as much flexibility as possible in terms of the kinds of materials students will be expected to use. 

Below, we provide a few examples of remote labs that used equipment students already had access to:
\begin{enumerate}
    \item \emph{``Students used their own computers/cellphones to acquire video data that they later analyzed using PASCO's Capstone software, so there was a requirement that they have access to a computer.''} [\intro, \notphyseng, \sm]
    
    \item Students used \emph{``the magnetic field sensors on their phones.''} [\intro, \physeng, \me, \eandm]
    
    \item \emph{``We used audacity on their laptop computers to analyze sounds.''} [\intro, \notphyseng, \sm, \waves]
    
    \item Building a pin-hole camera to observe the Sun.
\end{enumerate}
\vspace{10pt}

{\bf Sending equipment to students.} There was a general sense from instructors (and a desire from students, Figure \ref{fig:studentchallenges}) that finding some way to give students a hands-on experience was an essential part of the laboratory experience. Many instructors found success in mailing lab kits and equipment to students. However, this may be challenging for classes that have a large number of students, budgetary constraints, or do not want to increase fees for students. This is especially an important consideration for international students; one instructor pointed out, \emph{``Some students, due [to] international shipping constraints, cannot receive a kit. They will be sourcing the basic material themselves.''} [\intro, \notphyseng, \lrg] Another consideration is the availability of supplies --- with many courses across the country turning to remote lab instruction, ``off-the-shelf" lab kits such as the \href{https://www.iolabsinc.com/}{iOLab} or \href{https://esciencelabs.com/}{eScience} boxes may be in limited supply.  

A number of instructors chose to send Arduino micro-controller boards and basic electronics equipment to students. Some of these were choices made in the moment of transition, while others were part of ``Maker Lab'' courses that already used Arduinos in the classroom~\cite{Bradbury2020}. Simpler, and often cheaper, equipment may also provide the same experience. However, one must consider the health and safety (and liability and insurance) implications when sending equipment to students' homes. This is one possible advantage of commercial lab kits.

In Table~\ref{tab:remote-measurements}, we list the resources instructors reported using to send equipment to students. Below we enumerate some specific examples and comments on equipment that was mailed to students:
\begin{enumerate}
    \item \emph{``I mailed them printed off metersticks that could be mailed compactly and play-doh for the lens holders to be placed somewhat precisely along a meterstick.''}[\intro, \notphyseng, \sm, \optics]
    
    \item \emph{``Digital electronics seemed to be a pretty good platform for at-home experiments since the hardware is pretty robust and very inexpensive.''} [\bfy, \physeng, \sm, \electronics]
    
    \item \emph{``We use E-science instruction lab boxes sent to students. Boxes consist [of] very basic elementary objects to do simple labs. At first I was very skeptical, but it works very well.''} [\intro, \notphyseng, \sm]
    
    \item \emph{``We mailed each student 2 lenses and a diffraction grating and made the final 2 labs based on manipulating these components to study geometrical optics and diffraction. Students had to figure out how to mount components, how to use their phone as a light source, how to align and get images''} [\intro, \physeng, \me, \optics]
    
    \item \emph{``The students completed one lab to make a DC motor from a battery, paperclips, magnet, and wire.''} [\intro, \notphyseng, \sm, \electronics, \eandm]
    
    \item Some instructors suggested that they would have found \emph{``a hands-on device like IOLabs"} [\intro, \physeng, \lrg] helpful.  See Table~\ref{tab:remote-measurements} for more details of \href{http://www.iolab.science/}{iOLabs} and the recent paper by Leblond \textit{et al.} (2020)~\cite{Leblond2020}.

\end{enumerate}

\subsection{Analysis of data previously gathered by students}\label{sec:previously-gathered-data}
Similar to other lab activities (Section~\ref{sec:instructor-provided-data}) adopted in the remote setting, some courses shifted the focus to data analysis, but in this scenario using students' own data. \textit{``Instead of two projects, students extended their work on the first project, including many having to figure out issues with data collection without contact with apparatus.''} [\bfy, \physeng, \sm] This is an interesting activity in itself, as it indirectly teaches students the value of making good lab notes. In comparison to providing students with new data to analyze, this approach may address some aspects of student affect as the students have ownership over their data.

Often, this choice of activity coincided with extending the written aspect of the course (see Section~\ref{sec:written-comms}): \textit{``I had students analyze and report on previous measurements and focused on giving individual feedback on this written work.''} [\bfy, \physeng, \me] Some instructors took this as an opportunity to go further in developing skills associated with being a researcher: \textit{``For remote operation students wrote a PRL style article on an experiment they did in a previous quarter and engaged in a peer review exercise.''} [\bfy, \physeng, \sm]

\subsection{Physical equipment was controlled remotely}\label{sec:remote-controlled-equipment}
A number of survey respondents spoke of their desire to allow students to control lab equipment remotely. We provide a list of remotely-controlled labs in Table~\ref{tab:remote-measurements}. Remotely-controlled labs could be located at the instructor's own university or anywhere in the world. The short time available during the transition to remote labs meant that, in most cases, setting up remote access to in-house equipment was not feasible. However, some instructors did manage to do this:

\textit{``This was an advanced quantum optics lab. The equipment was housed in a lab at the university. Students logged into a PC via remote desktop. The optical arrangement was set up by the instructor. The computer controlled via USB various optical mounts (rotational and translational), plus piezo-electric.  The computer also connected to Arduino-based circuits/relays via USB to turn on/off equipment (lasers, detectors, beam blocker, LEDs) and FPGA circuits to process and record digital signals. Students observed the lab via webcams and connected with each other to do the lab via zoom. They had a span of a week to do the lab at any time they wanted. With coordination, the instructor was available for questions.''} [\bfy, \physeng, \sm, \quantum]

We have included this full quote in order to illustrate the amount of work needed to set up such equipment. Nevertheless, the motivation to do this work comes from wanting to provide students with the ability to perform their own measurements and to see the physics in action. This instructor found that \textit{``The student response [to the remote-controlled lab] was very positive.''} Other instructors who were able to set up in-house remote-controlled equipment commented on the benefits of that experience for students (e.g., working with LabView), but also noted that the process of setting up and maintaining the remote-controlled apparatus was frustrating and clumsy at times [\bfy, \physeng, \me]. 

In lieu of the experience and time required to do such a task, there exist a number of remote-controlled labs that are available online and were used by some instructors. These included the Princeton Plasma Physics Laboratory's remote glow discharge experiment, as well as the Universit{\"a}t der Bundeswehr M{\"u}nchen's Remotely Controlled Labs. In all of these remotely-controlled labs, the number of parameters available for students to vary is finite by construction, which makes the experience (in terms of the limited parameter space one can explore) similar to using simulations (see Section~\ref{sec:collect-data-from-sims}). 

A couple of instructors made use of the IBM Quantum Experience, which allows access to run quantum algorithms (and experiments) on their superconducting-qubit quantum computers. The website provides tutorials and a variety of interfaces to construct quantum algorithms. While this had a steep learning curve for both instructors and students, it was generally found to be successful in terms of learning outcomes: \textit{``I think the majority of students learned a significant amount of theory about quantum computing and acquired adequate skill in running remote quantum circuits on real quantum computers.''} [\bfy, \physeng, \sm, \quantum]

\subsection{Writing in labs}\label{sec:written-comms}
Communication skills, including scientific documentation and writing, are often included as learning goals for physics lab classes~\cite{KozminskiEtal2014}. Instructors may have a variety of goals for incorporating writing in lab classes---from helping students develop content mastery to having students engage in realistic scientific practices such as argumentation or peer review~\cite{HoehnLewandowski2020}. Compared to developing technical or hands-on skills, writing is one of the important aspects of lab classes that can more easily be maintained remotely. Survey respondents reported utilizing most of the same writing assignments in the remote version of their course as compared to in person, with a decrease in the number of people using lab notebooks and an increase in the number of people having students read scientific papers and write a literature review. 

In the transition to remote teaching, some instructors took the opportunity to place a heavier emphasis on writing. For example, one instructor included a project proposal where students had \emph{``to do research on some type of user facility or instrument and come up with an experimental proposal; this involves doing a literature review and working with [the instructor] to refine experimental designs and parameters.''} [\bfy, \physeng, \sm] Though not ideal as a complete replacement for hands on experimentation, this is one way that writing can be used to address some of the key elements of a lab class, particularly for student-designed projects in advanced labs, and it worked well as an immediate solution to the challenge of creating a remote lab class. Other instructors maintained the same writing assignments, but chose to emphasize them through a modified grading scheme.

Other instructors had students write about prior experiments they had conducted or data they had collected in-person before the transition to remote teaching. In one example, \emph{``students wrote a PRL style article on an experiment they did in a previous quarter and engaged in a peer review exercise.''} [\bfy, \physeng, \sm] Another instructor used writing to address goals of the class because \emph{``Even though no lab work occurred after remote instruction began, students had to rely on their notebooks and previous data collection to complete required oral presentations and written reports, both considered part of `lab skills.' ''}[\bfy, \physeng, \sm] This is in line with recommendations from Stanley and Lewandowski~\cite{Stanley2018} for using notebooks in upper-division lab classes in a way that promotes authentic documentation by requiring students to rely on their own (or others') notebooks. 

Though some instructors stopped using lab notebooks after the transition to remote teaching, others switched from hard copy to electronic lab notebooks (ELNs), utilizing tools like \href{https://www.labarchives.com/}{LabArchives} or \href{https://www.google.com/docs/about/}{Google Docs}. In one example of an intro class, the instructor reported that students were more engaged with the LabArchives ELNs compared to the in person paper notebooks: some students tended to write more during the lab activities and they appreciated being able to easily include graphs/diagrams as well as having access to the ELN at any time. The instructor said that because \emph{``the students gave positive feedback on that...I'm considering switching to e-notebooks next year.''} [\intro, \physeng, \sm] Other instructors appreciated the grading ease of ELNs, saying \emph{``I had resisted electronic lab notebooks for years.  Now, I was forced to try it out.  It seemed to go just fine, and it was easier to grade (as opposed to lugging around a pile of notebooks).''} [\intro, \notphyseng, \me] These, and other, benefits of ELNs have been previously documented in the literature~\cite{Eblen-Zayas2015}.

Some instructors replaced written lab reports with other media like video presentations. For example, one instructor said that students would \emph{``turn in their last lab as a video recording of them describing their procedure, data and analysis, and results/conclusions. The video will show their data, graphs, and written work, recorded along with narration on their cell phone.''} [\intro, \notphyseng, \me] In other cases, instructors supplemented traditional forms of writing (e.g., reports, notebooks) with other types of writing assignments. In one advanced lab class, student-designed final projects culminated in both a lab report and a blog post, in which the students had to describe their experiment in more informal or colloquial terms. The blog post assignment replaced the typical oral presentations as something that could easily be done asynchronously. The goal of the blog post assignment was to have students practice writing about experimental physics for different audiences; students found it to be a fun and useful exercise. [\bfy, \physeng, \sm]

\section{Student agency and engagement}\label{sec:student-agency}
\begin{figure}[h]
    \centering
    \includegraphics[scale=0.6]{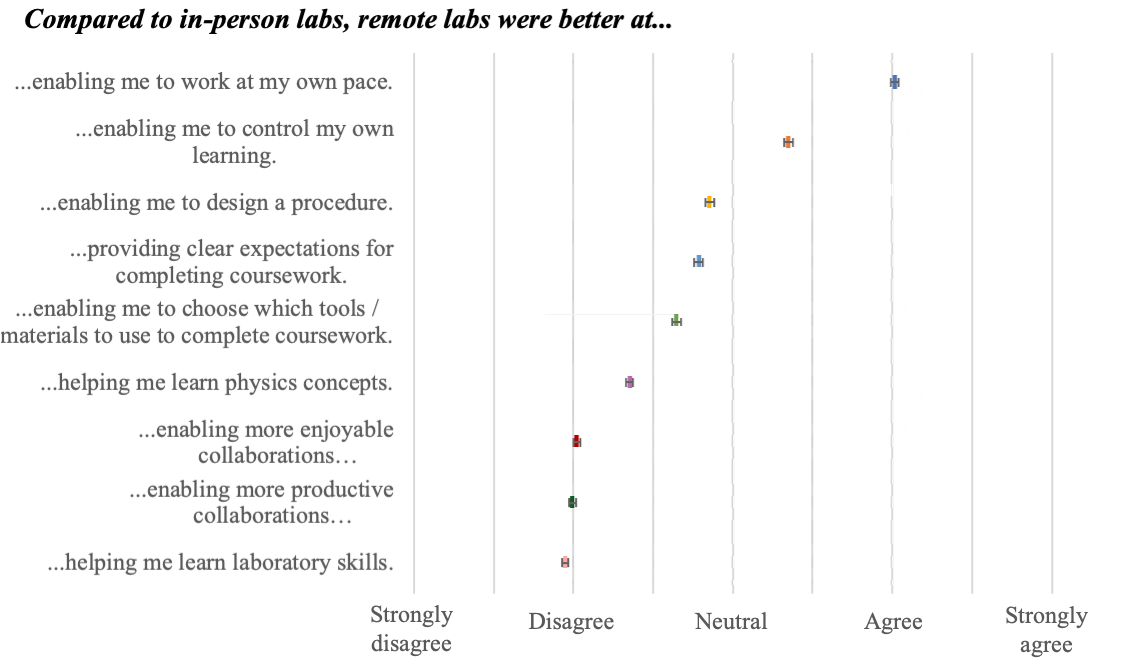}
    \caption{Students were asked, ``Compared to in-person labs, remote labs were better at...'' and then responded to the following statements with their level of agreement. We show the mean response from approximately 2200 students. The error bars represent the standard error of the mean. We calculated the mean by assigning a response of ``Strongly disagree" = 0, ``Disagree" = 1, and ``Neutral" = 2, ``Agree = 3", and ``Strongly agree = 4".}
    \label{fig:inpersonlabs}
\end{figure}

One benefit of remote classes is that they can provide more opportunities for student agency. For example, many students felt that remote labs were better at enabling them to work at their own pace and to control their own learning (Figure \ref{fig:inpersonlabs}).

When it came to designing their own procedures, agency in tool/material choices, and learning concepts and skills, a majority of the students felt that the remote classes were the same or worse than in-person labs. Similarly, many instructors expressed challenges with maintaining student agency and engagement in the remote setting:

\begin{enumerate}
    \item \emph{``The other big problem was student engagement. Without setting up structures from the get-go, it was too easy for students to just drift.''} [\bfy, \physeng, \sm, \electronics]

    \item Another challenge was \emph{``having students think about the online experience with the same intensity they considered in-person labs.''} [\intro, \physeng, \sm]
    
    \item \emph{``As soon as pass/fail grading was announced, some groups stopped turning in lab reports.''} [\bfy, \physeng, \me, \quantum]
\end{enumerate}

However, a few instructors who had open-ended labs found much success: \emph{``The labs that worked best were the more open-ended when students used a PhET simulation to answer a question of their own choosing.''} [\intro, \physeng, \sm, \eandm, \optics] Some instructors took this a step further and transformed the remote course to work on open-ended ``research like projects'' compared to ``cookbook'' labs before the remote transition. One instructor commented, \emph{``The level of student engagement was much higher in the remote format. Students were much more engaged in problem solving and making meaningful decisions about what to do and how to do it.'' [\bfy, \physeng, \me]}

\section{Collaboration and interactions}\label{sec:communication}
\subsection{Group work}\label{sec:group-work}
\begin{figure}[h]
    \centering
    \includegraphics[scale=0.9]{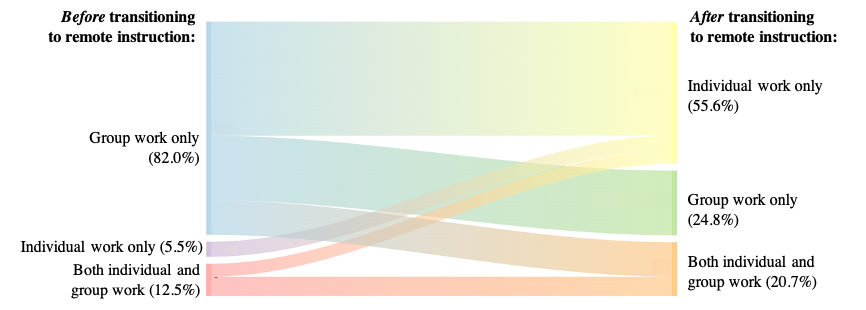}
    \caption{The Sankey plot shows the change in the nature of interaction students took part in --- individual, group work, or a combination --- from before (left side of the plot) and to after (right side of the plot) the transition to remote instruction for the courses represented in the instructor survey. The lines show the proportion of courses that either stayed the same or changed from one mode of student interaction to another during the transition. The width of each line is proportional to the number of instructors who reported that type of transition.}
    \label{fig:groupwork}
\end{figure}

After the switch to remote instruction, most classes moved to individual work and incorporated less group work (Figure \ref{fig:groupwork}). We also see in Figures \ref{fig:studentchallenges} and \ref{fig:inpersonlabs} that many students felt that they did not have as productive nor enjoyable collaborations after they switched to remote labs and expressed that having a partner/group to help conduct experiments was one of the greatest challenges.

It was easiest in the rapid transition (and, in some cases, most equitable) to have students work primarily individually, especially when students were spread across different time zones. However, given that social interactions and collaboration are paramount to learning and doing science, we have a few recommendations and successful examples of how to get students to engage in group work:

\begin{enumerate}
    \item Use Zoom breakout rooms feature: \emph{``[In Zoom we had] individual breakout rooms to preserve small group learning environment where students develop the lab and challenge each others' ideas. This also preserved the ability of the TA to give meaningful as needed scaffolding to the students as they would in the regular classroom.''} [\intro, \physeng, \sm]

    \item Keep groups small: \emph{``It was actually less of a problem for them to collaborate than I expected, as long as I kept the groups to three students.''} [\intro, \notphyseng, \me]

    \item Students don't necessarily have to have a good internet speed/connection to engage in a group discussions. Discussion boards on collaborative software such as your school's learning management systems can be used to foster (synchronous and asynchronous) discussions. \href{https://slack.com/}{Slack}, or other similar tools, can essentially act as a chat room for your entire class. It has workspaces that allow you to organize communications by channels for group discussions and allows for private messages to share information, files, and more all in one place. 
    
    \item \href{https://colab.research.google.com/}{Google Colab}, \href{https://jupyter.org/}{Jupyter Notebook}, and \href{https://github.com/}{GitHub} (see Table~\ref{tab:teams}) have features that allow for collaborative coding and making notes.
\end{enumerate}

\subsection{Asynchronous and synchronous class sessions}

We include a section on benefits and challenges of asynchronous versus synchronous class activities in this report for two reasons: (1) Lab courses rely heavily on group work and collaboration; therefore, considerations of equity (family situations/schedules, access to stable internet, time zones, etc.) and building/maintaining community is an even more challenging balance than for most traditional lecture courses; (2) Approximately 50\% of the instructors we surveyed responded that if they were to teach this course remotely again, they would ``structure course time differently (e.g., synchronous vs. asynchronous)".

Finding the right balance between asynchronous and synchronous class sessions will be context dependent (e.g., in a very small class, you can check in on the situation of individual students and even ask them what they prefer, but in a large class, it is not recommended to employ exclusively synchronous activities) and based on institutional requirements. Many instructors cited trying to get the best of both worlds by recording synchronous lectures for students who could not participate. This may be an easy and quick solution; however, it brings up other equity issues of having two different experiences within the same class. Below, we provide some examples of how instructors implemented synchronous and asynchronous labs.

\subsubsection{Synchronous}\label{sec:synchronous}

While there are challenges around how to conduct group discussions in a synchronous online format, one-on-one meetings to discuss lab projects have generally been successful: \emph{``I really like that when I talk to students individually we get to have the types of conversations we would in an in-person class.''} [\bfy, \physeng, \sm]

A number of instructors did live labs via videoconferencing, where the students watched the instructor take data, which they then analyzed. Some instructors took this a step further by having students guide them as they conducted the lab.

Common issues with synchronous labs were low attendance, equity issues, and video quality. However, one instructor pointed out that \emph{``going synchronous makes life much more easy for the teacher than providing high quality videos.''} [\bfy, \physeng, \me]

Some benefits of synchronous labs were that they allowed for group work (especially if the groups were small, which can be facilitated through breakout rooms) and promoted community and accountability. Not only can synchronous labs be an opportunity for students to work collaboratively, but also for students to engage with a teaching assistant: \emph{``Students were invited to attend their usual lab time on Zoom to discuss together and/or with their TA.''} [\intro, \physeng, \lrg] However, another instructor warned that they really needed to \emph{``train our TAs to handle the labs in that way [remotely].''}  [\intro, \physeng, \me]

Lastly, synchronous meetings can be a way to ``check-in'' and connect with students beyond the class, especially during the the pandemic. One instructor expressed that they used Zoom to \emph{``maintain the weekly updates... [which] not only allowed me to get a status report on projects also monitor mental health of students.''} [\bfy, \physeng, \sm]

\subsubsection{Asynchronous}\label{sec:asynchronous}

Asynchronous instruction has a number of advantages:
\begin{enumerate}
    \item Acknowledges and caters to a variety of personal situations (for students \emph{and} instructors);
    \item Potentially good for student agency, as it allows students to do work at their own pace and on their own schedule (see Figure \ref{fig:inpersonlabs});
    \item Works well when one does not need to have interaction with students (i.e., lecture or lab introduction).
\end{enumerate}

Personal factors might be the most motivating reason to use asynchronous teaching methods. For example, one instructors said \emph{``many of my students had to take on additional responsibilities at home, so I had to make sure that the labs could be done individually so that students could do them asynchronously.''} [\intro, \notphyseng, \sm] The success of delivering asynchronous course material required a level of planning and consideration on behalf of some instructors: \emph{``Students needed time to adjust to the quick transition, by going asynchronous and having very detailed, step by step instructions, students could make this transition at their own pace.''} [\intro, \stem, \me]

Not only were considerations of student situations being made, but instructors had to account for their own home lives too. This motivated one instructor \emph{``to [do] things asynchronously because I was home with two small children.''} [\intro, \notphyseng, \sm] Nevertheless, effective lab courses can still be achieved with high quality videos like those from Pivot Interactive: \emph{``Their collection of videos is really good.''} [\intro, \notphyseng, \sm] Or if they are paired with additional student activities that increase student agency, such as conducting authentic research (Section \ref{sec:big-data}), sending students equipment (Section \ref{sec:student-collected-data-at-home}), or having students design their own procedures (Section \ref{sec:written-comms}).

\section{Looking towards next semester}
As we look toward the Fall 2020 term, many universities plan to have hybrid models that consist of both remote and in-person portions of the courses. However, most universities are allowing students to opt-in to a completely remote experience at any point and additionally, have warned instructors to prepare to rapidly switch to completely remote if the school needs to close down again. The hybrid model opens many different opportunities for lab courses that were not described in this report. Some faculty plan to front load the more technical labs in the beginning of the semester and have modeling/computation based labs toward the end; others have suggested they will rotate the students who attend the lab in-person each week. We again encourage thinking about equity when designing these hybrid courses such that students who choose to take the class remotely (or need to for health, family, or other reasons) have an experience that is equally considered as the in-person component. 

We hope to continue collecting data on student and instructor experiences teaching in the Fall 2020 term. We encourage instructors interested in evaluating the effectiveness of their lab and the remote experience to survey their students at the beginning and end of the semester. Our research group has developed \href{https://jila.colorado.edu/lewandowski/research/e-class-colorado-learning-attitudes-about-science-survey-experimental-physics}{The Colorado Learning Attitudes about Science Survey for Experimental Physics (E-CLASS)}, a broadly applicable assessment tool for undergraduate physics lab courses that assesses students views about their strategies, habits of mind, and attitudes when doing experiments in lab classes. E-CLASS has been adapted to include supplemental questions about remote/hybrid lab experiences to help instructors reflect on their own strategies and help inform the larger community about remote experiences that students found most successful. Instructors can sign-up to administer E-CLASS to their students by filling out the form on the E-CLASS website (linked above).

\section{Conclusions}\label{sec:next-semester}

Despite the seemingly insurmountable challenges many faced last term, physics lab instructors rose to the occasion and employed a variety of creative approaches and strategies in order to provide opportunities for students to access ``lab-like'' learning online. For some instructors, the move to remote/hybrid teaching may be a unique opportunity to transform the lab course---rethinking learning goals, implementing course-based undergraduate research experiences (CUREs), having at-home maker spaces or labs that focus heavily on experimental design and modeling to increase student agency, or completely restructuring both the lectures and labs to have investigative science learning environments (ISLEs).  

We encourage the reader to consider some of the larger themes that emerged while compiling these data:
\vspace{2mm}
\begin{enumerate}
    \item Be prepared to deal with technical issues, from internet connection problems to access to resources, especially if planning for students to conduct labs at home.
    \item The flexibility provided by open-ended projects, if managed successfully, work well in the remote environment.
    \item Synchronous, short meetings with small groups via videoconferencing anecdotally worked better than longer meetings with larger groups.
    \item Do not assume that all students have access to internet and household materials.
    \item When deciding which materials or technological tools to utilize in a remote class, consider the accessibility for students with cognitive or physical disabilities. 
    \item Recordings of synchronous meetings can be made available to students to ensure access to course material.
    \item Both preparation time for instructors and coursework time for students can be dramatically increased when doing the course remotely. Keep this in mind when planning a remote lab course to avoid overwhelming students (and instructors) with work. 
    \item This was, and still is, a new situation for everyone, so things will go wrong---that is okay.
\end{enumerate}

\vspace{2mm}

\comment{
Here, we summarize a few specific, practical things to consider:
\vspace{2mm}
\begin{enumerate}
    \item For group work, especially if measurements are being made using household equipment, requiring only one student to perform measurements while others take on different roles in the group (planning, data analysis, writing, etc.) is a possible way to provide an equitable experience for students.
    \item TAs will need to be provided training on how to conduct teaching in the remote setting.
    \item Group work with students in different time zones may prove challenging. One should make sure that assembled groups will be able to effectively collaborate.
\end{enumerate}
\vspace{2mm}
}

As we conclude this report, we reiterate that there are many metrics of success that one might apply to a remote lab class during this time of transition and uncertainty. Ensuring that all students have access to learning opportunities, making it through without a disaster, and achieving specific learning outcomes are all reason to celebrate and feel proud of responding to the challenge of teaching lab classes remotely. Additionally, access to technology, having a quite space to work, family responsibilities, and both mental and physical health are not only challenges for our students, but also for instructors. Whether trying to simply making it through the upcoming term as painlessly as possible or using it as an opportunity to transform the course, we hope this report has provided some inspiration for curricular and pedagogical strategies that will enable instructors to meet their learning goals and engage their students in physics laboratory learning an equitable way. 

Lastly, we, as well as many instructors, believe that remote teaching of labs should be temporary, and, when health and safety conditions allow, should be moved back to in-person instruction. Although instructors have gone to great lengths to give students the best possible learning experiences under severe constraints, many critical learning goals are hard, if not impossible, to meet in a fully remote class. We look forward to welcoming our students back to in-person classes where they can have the opportunity to participate in the full process of experimental physics.

\addcontentsline{toc}{section}{Acknowledgments}
\section*{Acknowledgments}

We would like to thank Benjamin Pollard, Mary-Ellen Philips, and Joe Wilson for their contributions to this work, and all the instructors and students who shared their experiences with us. This work is supported by NSF RAPID Grant (DUE-2027582).

\clearpage
\addcontentsline{toc}{section}{References}
\bibliographystyle{unsrt}
\bibliography{rapid-report}

\addcontentsline{toc}{section}{Index}\label{sec:index}
\printindex

\newpage

\appendix
\setcounter{table}{0}
\renewcommand\thetable{A\arabic{table}}
\section{Technological Resources}\label{sec:technology}
The variety of different technological resources that are available can be overwhelming. In this Appendix, we tabulate the resources that instructors reported using in their courses. We also include other resources that the authors are aware of, noting that these are not exhaustive lists. We do not comment here on whether a specific technology was effective at the job it was designed for, as the efficacy of any technology depends on the course goals, content, instructor experience, and institutional requirements among numerous other factors.

\begin{table}[h]
    \centering
    \caption{Simulation tools that were mentioned by instructors in our survey (in alphabetical order). See Section~\ref{sec:collect-data-from-sims} for a discussion and examples of the use of some of these simulations.}
    \label{tab:simulations}
    \vspace{5pt}
    \begin{tabular}{L{4cm}lL{9cm}}
    {\large \textbf{Simulation}} & {\large Model} & {\large Description} \\ \hline\hline
     \href{https://bridgedesigner.org/}{Bridge designer 2016}& Free & Students apply engineering design skills and physics knowledge to design a bridge; simulation of forces and loads on the bridge structure. \\
     
     \href{https://fritzing.org/home/}{Fritzing} & Free & An open-source CAD design tool for electronic circuit boards. Has the ability to manufacture printed circuit boards. \\
     
     \href{https://virtualphysicslabs.ket.org/}{KET} & Paid & Virtual physics labs including teaching materials. \\
     
     \href{https://matterandinteractions.org/student/}{Matter \& Interactions} & Free & Interactive demos on Mechanics and Electric \& Magnetic interactions written in VPython and runs through a web browser. \\
     
     \href{https://www.multisim.com/}{Multisim Live} & Free & Online circuit simulator\\
     \href{https://www.compadre.org/physlets/}{Physlets}  & Free & ``Interactive illustrations, explorations, and problems for introductory physics''\\
     
     \href{https://www.compadre.org/osp/}{Open Source Physics} & Free & Compilation of Java simulations, student coding resources, and tracking software for video analysis.  \\
     
     \href{https://openstax.org/subjects/science}{OpenStax} & Free & Open source textbook on physics (with embedded PhET simulations). \\
     
     \href{http://ophysics.com/}{oPhysics} & Free & Interactive simulations of phenomena including kinematics, forces, conservation, waves, light, E\&M, rotation, fluids, and modern physics. Uses the \href{https://www.geogebra.org/}{Geogebra} software, which has its own compilation of simulations.\\
     
     \href{https://phet.colorado.edu/}{PhET} & Free & Physics, Math, and other science simulations in HTML5, Flash, and Java. Resources and advice for using as remote teaching tools \href{https://docs.google.com/document/u/1/d/e/2PACX-1vQMwNbLNOtwTdS4sWbXx4dXnJHDoENTgyvVl4Vwrq6NbC3ijlCrpPncSTItitHFRv9mp6-FFCL2uYp8/pub}{available on their website}.\\
     
     \href{https://www.pivotinteractives.com/}{Pivot Interactives}& Paid & Videos of real lab experiments overlaid with virtual measurement devices allowing students to perform measurements themselves. Videos for a large variety of different parameters allow students to explore the experiment. Includes worksheets.\\
     
     \href{https://www.physport.org/curricula/quilts/}{Quantum Interactive Learning Tutorials (QuILT)} & Free & Packaged material for teaching quantum mechanics including Java, PhET, and Open Source Physics simulations. \\
     
     \href{https://www.mathworks.com/products/simscape.html}{Simscape} & Paid & Model and simulate multi-domain physical systems in the MathWorks Simulink environment based on MATLAB.\\
     
     \href{http://www.physics.pomona.edu/sixideas/resources.html}{Six Ideas That Shaped Physics} & Free & Simulation resources to coincide with chapters from the textbook. \\
     
     \href{http://bwrcs.eecs.berkeley.edu/Classes/IcBook/SPICE/}{SPICE} & Free & Open-source analog circuit simulator (some proprietary versions exist - PSPICE and HSPICE). \\ 
     
     \href{http://thephysicsaviary.com/}{The Physics Aviary} & Free & A set of physics simulations and associated resources.\\
     
     \hline
    \end{tabular}
\end{table}     
     
\begin{table}[h]
    \centering
    \caption{Resources for students to perform measurements outside of the laboratory.}
    \label{tab:remote-measurements}
    \vspace{5pt}
    \begin{tabular}{L{4cm}lL{9cm}}
    {\large Resource} & {\large Model} & {\large Description} \\ \hline\hline     
     \textbf{Smartphone apps} & & Section~\ref{sec:student-collected-data-at-home}\\ \hline
     \href{https://phyphox.org/}{Phyphox} & Free & Collects (and processes) data from smartphone sensors depending on the device (accelerometers, rotation, light intensity, magnetic field, GPS location, audio, pressure). Allows for connecting to a computer using a web browser to run experiments and transfer data. \\
     \href{https://sciencejournal.withgoogle.com/}{Google Science Journal} & Free & Collects data from smartphone sensors (similar to Phyphox). Includes integration with Google Drive and website includes some activities for teachers. \\
     \hline
     \textbf{Sending equipment} & & Section~\ref{sec:student-collected-data-at-home}\\ \hline
     \href{https://www.arduino.cc/}{Arduino} & Paid & A variety of microcontrollers and kits that can be used for digital and analog programming and sensing. A lot of resources are available around Maker labs~\cite{Bradbury2020}.\\
     \href{https://www.raspberrypi.org/}{Raspberry Pi} & Paid & Similar to Arduinos, runs linux and can control and run sensors. Requires extra interfaces to handle analog inputs. \\
     \href{https://esciencelabs.com/}{eScience lab boxes} & Paid & Commercial provider of lab kits for remote courses. \\
     \href{https://www.digikey.com/}{Digikey} & Paid & The ``Bill of Materials'' manager was used by one instructor ``to drop-ship items out to students inexpensively and quickly.'' \\
     \href{http://www.iolab.science/}{iOLab} & Paid & Numerous sensors (force, acceleration, velocity, displacement, magnetic field, rotation, light, sound, temperature, pressure, and voltages down to a few $\mu$V) combined into a single device that can be sent to students. Data is transferred and analyzed using computer software. For using iOLab with remote teaching see the recent paper by Leblond \& Hicks~\cite{Leblond2020}. \\ \hline
     
     \textbf{Remote control of lab equipment} & & Section~\ref{sec:remote-controlled-equipment} \\ \hline
     \href{http://stem.open.ac.uk/study/openstem-labs}{OpenSTEM Labs} & Paid & Remotely-controlled labs run by the Open University for their own students. \\
     \href{https://quantum-computing.ibm.com/}{IBM Quantum Experience} & Free & Access IBM's quantum computers to run quantum algorithms. Includes tutorials and documentation. \\
     \href{https://www.pppl.gov/RGDX}{PPPL remote glow discharge experiment} & Free & Remote access to the Princeton Plasma Physics Lab experiment designed for students to learn about plasma. \\ 
     \href{http://rcl-munich.informatik.unibw-muenchen.de/}{Remotely Controlled Labs} & Free & Access to labs provided by the Universität der Bundeswehr München. Labs on electron diffraction, Millikan's experiment, optical computed tomography, speed of light, world pendulum, oscilloscope, photoelectric effect, semiconductor characteristics, wind tunnel, optical Fourier transformation, and diffraction and interference.\\
     
     \hline
    \end{tabular}
\end{table}

\begin{table}[h]
    \centering
    \caption{Resources for working in teams remotely.}
    \label{tab:teams}
    \vspace{5pt}
    \begin{tabular}{L{4cm}lL{9cm}}
    {\large Resource} & {\large Model} & {\large Description} \\ \hline\hline       
     \textbf{Coding collaboratively} & & \\ \hline
     \href{https://colab.research.google.com/}{Google Colabs} & Free & Online collaborative code site using Python with free remote processing. \\
     \href{https://jupyter.org/}{Jupyter notebooks} & Free/Paid & Open-source online and local code notebooks in Python, C++, Julia, R, and Ruby. If wish to host a Jupyterhub to run student codes (so that they do not have to rely on their own hardware) a paid hosting option exists.\\      \href{https://github.com/}{GitHub} & Free/Paid & Web-based graphical interface for a Git repository that provides access control and several collaboration features, such as a wikis and basic task management tools for coding project. \\ \hline
     
     \textbf{Virtual lab notebooks} & & Section~\ref{sec:written-comms} \\\hline
     \href{https://www.google.com/docs/about/}{Google Docs} & Free & Word processing tool allowing multiple users to edit simultaneously.\\
     \href{https://www.microsoft.com/en-us/microsoft-365/onenote/digital-note-taking-app?rtc=1}{Microsoft OneNote} & Paid & Note keeping, organization, and collaboration tool. Included with most institutional Microsoft Office licenses.\\
     \href{https://www.labarchives.com/}{LabArchives} & Paid & Professional digital lab notebook tool. Education version includes customizeable course packs with pre-written labs. Was made available for free during the pandemic.\\
     
     \hline
    \end{tabular}
\end{table}

\end{document}